\documentclass[12pt]{article}

\usepackage{amsmath}
\usepackage{amsthm}

\theoremstyle{remark}

\begin{document}

{\noindent\bf\Large The Complex Symmetry Gravitational Theory as a
New Alternative of Dark Energy}\\
\\
{\bf Ying Shao $^1$ Yuan-xing Gui \footnote{Department of Physics,
Dalian University of Technology, Liaoning Province 116023, China;
\\e-mail : thphys@dlut.edu.cn} and Wei Wang
$^1$}\\
\\
\leftskip 1.5 true cm
\parbox{12cm}
{\small{We propose that complex symmetry gravitational theory
(CSGT) explain the accelerating expansion of universe. In this
paper universe is taken as the double complex symmetric space.
Cosmological solution is obtained within CSGT. The conditions of
the accelerating expansion of universe are discussed within CSGT.
Moreover, the range of equation of state of matter
$\omega_\varepsilon$ is given in the hyperbolic imaginary space.
 }}\\

\leftskip 0 true cm

\section{INTRODUCTION}
One of the most surprising discoveries of the past decade is that
expansion of the universe is currently speeding up rather than
slowing down [1]. The case receives support from CMB [2] and the
observations of supernovae and gravitational clustering [3,4]. The
accelerating expansion has been attributed to dark energy by which
the universe is dominated. But, there is no direct evidence of
dark energy and the nature of dark energy remains mysterious. At
present, several candidate for dark energy have been presented.
The simplest and most obvious candidate is cosmological constant
[5]. Although cosmological constant appears to satisfy all
observations, the fine-tuning
 difficulties have prompted theorists to investigate a variety of alternative models where equation of state of dark energy
 is time dependent. Popular dark energy models include Quintessence [6],Braneworld models [7],Chaplygin gas [8],
 Phantom energy [9] and modified gravity [10,11]. Currently J.W.Moffat has proprosed that the nonsymmetric gravitational
 theory(NGT)[12,13,14] explain the accelerating expansion of universe [15], which is a new attempt. NGT and complex symmetric
 gravitational theory(CSGT)[16,17] are also generalized gravitational theory presented as the unified field theory of gravity
 and electromagnetism.\\
 \mbox{}\hspace{20pt}In the following, we shall apply CSGT to explain the accelerating expansion of universe. In this theory,
 universe is the double complex symmetry space. Our paper is organized as follows. We briefly review the
foundations of CSGT in next section. In Sec III, the metric takes
the canoncial Gaussian form for comoving coordinates. We solve
Einstein's field equation without cosmological constant and obtain
the cosmological solution. In Sec IV, we
 investigate that when satisfying the some condition, matter in hyperbolic imaginary space gives rise to the accelerating expansion of
 universe. Furthermore, we give the range of equation of state of matter $\omega_\varepsilon$
 in hyperbolic imaginary space. In Sec V, we formulate our conclusions.
\section{THE FOUNDATION OF CSGT}
In complex symmetry gravitational theory (CSGT), metric tensor is
a complex symmetric tensor. Correspondingly, connection and
curvature are forced to be complex. The real diffeomorphism
symmetry of standard Riemannian geometry is extended to complex
diffeomorphism
symmetry.\\
\mbox{}\hspace{20pt}We consider choosing an complex manifold of
coordinates ${\cal M}^4_C$ and an complex symmetric metric defined
by [18,19,20]
$$
{g_{\mu\nu}}={s_{\mu\nu}}+J{a_{\mu\nu}},\eqno(1)
$$
where ${s_{\mu\nu}}$ and ${a_{\mu\nu}}$ are the real symmetric
tensors, the double imaginary unit$J=i,\varepsilon$. The real
contravariant tensor ${s^{\mu\nu}}$ is associated with
${s_{\mu\nu}}$ by the relation
$$
{s^{\mu\nu}}{s_{\mu\sigma}}={\delta^\nu_\sigma},\eqno(2)
$$
and also
$$
{g^{\mu\nu}}{g_{\mu\sigma}}={\delta^\nu_\sigma},\eqno(3)
$$
With the complex spacetime is also associated a complex symmetric
connection
$$
{\Gamma}_{\mu\nu}^{\lambda}={\Delta}_{\mu\nu}^{\lambda}+J{\Omega}_{\mu\nu}^{\lambda},\eqno(4)
$$
where ${\Delta}_{\mu\nu}^{\lambda}$ and
${\Omega}_{\mu\nu}^{\lambda}$ are also the real symmetric tensors.
The complex symmetric connection $ {\Gamma}_{\mu\nu}^{\lambda}$ is
determined by the equations
$$
{g_{\mu\nu;\lambda}}={\partial_\lambda}{g_{\mu\nu}}-{g_{\rho\nu}}{g_{\mu\lambda}^\rho}-{g_{\mu\rho}}{\Gamma_{\nu\lambda}^\rho}=0,\eqno(5)
$$
Furthermore,we obtain the generalized curvature tensor
$$
R_{\mu\nu\sigma}^{\lambda}=-{\partial_\sigma}{{\Gamma}_{\mu\nu}^{\lambda}}+
{\partial_\nu}{{\Gamma}_{\mu\sigma}^{\lambda}}+{{\Gamma}_{\rho\nu}^{\lambda}}{{\Gamma}_{\mu\sigma}^{\rho}}
-{{\Gamma}_{\rho\sigma}^{\lambda}}{{\Gamma}_{\mu\nu}^{\rho}},\eqno(6)
$$
and a contracted curvature tensor
$$
{R_{\mu\nu}}:=R_{\mu\nu\sigma}^{\sigma}={Q_{\mu\nu}}+J{P_{\mu\nu}},\eqno(7)
$$
where $Q_{\mu\nu}$ and $P_{\mu\nu}$ are the real symmetric
tensors. From curvature tensor,we can obtain the four complex
Bianchi identities
$$
{{\bigg(}{R^{\mu\nu}}-\frac{1}{2}g^{\mu\nu}R{\bigg)};_\nu}=0,\eqno(8)
$$
\mbox{}\hspace{20pt}The CSGT action is denoted by [16,18]
$$
S={S_{grav}}+{S_M},\eqno(9)
$$
where $S_{grav}$ and $S_M$ are gravity action and matter action
respectively.\\
\mbox{}\hspace{20pt}We choose a following real action to guarantee
a consistent set of field equations
$$
{S_{grav}} =\frac{1}{2}\int{d^4}x{\bigg[}{{\cal
 G}^{\mu\nu}}{R_{\mu\nu}}+{{\big(}{{\cal
 G}^{\mu\nu}}{R_{\mu\nu}}{\big)}^\dag}{\bigg]},\eqno(10)
$$
and the matter part of action is
$$ \frac{1}{\sqrt{-g}}(\frac{\delta
S_M}{\delta g^{\mu\nu}})=8\pi G{T_{\mu\nu}},\eqno(11)
$$
where ${\cal G}^{\mu\nu}:=\sqrt{-g}g^{\mu\nu}={\cal
 S}^{\mu\nu}+J{\cal
  A}^{\mu \nu}$£¬``$\dag$''denotes complex
 conjugation,$T_{\mu\nu}=\tau_{\mu\nu}+J{\tau_{\mu\nu}^{'}}$ is a
 complex symmetric source tensor. The variation with respect to $g^{\mu\nu}$ yields the field
 equations
 $$
 {\cal R}_{\mu\nu}-\frac{1}{2}{g_{\mu\nu}}{\cal R}=-8\pi
 G{{\cal T}_{\mu\nu}},\eqno(12)
 $$
 where ${{\cal R}_{\mu\nu}}=\sqrt{-g}{R_{\mu\nu}}$,${\cal R}={\cal
 G}^{\mu\nu}{R_{\mu\nu}}$,${{\cal
 T}_{\mu\nu}}=\sqrt{-g}{T_{\mu\nu}}$. Dividing the above equation by
 $\sqrt{-g}$, complex field equations (12) are
$$
R_{\mu\nu}-\frac{1}{2}{g_{\mu\nu}}R=-8\pi
 G{T_{\mu\nu}},\eqno(13)
 $$
Eq.(13) is written as
$$
R_{\mu\nu}=-8\pi
G({T_{\mu\nu}}-\frac{1}{2}{g_{\mu\nu}}T),\eqno(14)
$$
\section{COSMOLOGICAL SOLUTION AND MODIFIED FRIEDMANN EQUATION}
Let us consider a real line element
$$
{dS^2}=-{dt^2}+\alpha(r,t){dr^2}+\eta(r,t)({d\theta^2}+{\sin^2}\theta{d\phi^2}),\eqno(15)
$$
where $\alpha(r,t)$ and $\eta(r,t)$ are functions of real $r$ and
$t$. The complex symmetric tensor $g_{\mu\nu}$ is determined by
$$
{g_{00}}(r,t)=-(1+J),
$$
$$
{g_{11}}(r,t)=\mu(r,t)=\alpha(r,t)+J\beta(r,t),
$$
$$
{g_{22}}(r,t)=\gamma(r,t)=\eta(r,t)+J\xi(r,t),
$$
$$
{g_{33}}(r,t)=\gamma(r,t){{\sin^2}\theta}=[{\eta(r,t)+J\xi(r,t)}]{\sin^2}\theta,\eqno(16)
$$
Solving the $\Gamma^\lambda_{\mu\nu}$ and substituting into
Eq.(6), we get
$$
R_{00}=\frac{\mu_{tt}}{2\mu}-\frac{{\mu_t}^2}{4{\mu}^2}+\frac{\gamma_{tt}}{\gamma}-\frac{{\gamma_t}^2}{2\gamma^2},\eqno(17)
$$
$$
R_{11}=-\frac{\mu_{tt}}{2}+\frac{\mu_t^2}{4\mu}+\frac{\gamma_{rr}}{\gamma}-\frac{{\gamma_t}{\mu_t}}{2\gamma}-\frac{{\mu_r}{\gamma_r}}
{2\mu\gamma}-\frac{\gamma_r^2}{2\gamma^2},\eqno(18)
$$
$$
R_{01}=\frac{\gamma_{tr}}{\gamma}-\frac{{\mu_t}{\gamma_r}}{2\mu\gamma}-\frac{{\gamma_t}{\gamma_r}}{2\gamma^2},\eqno(19)
$$
where subscripts mean derivative with respect to
$t$ and $r$ respectively.\\
\mbox{}\hspace{20pt}In complex spacetime,the energy-momentum
tensor takes
$$
T^{\mu\nu}=[({\rho_C}+{p_C}){U^\mu}{U^\nu}+{p_C}{g^{\mu\nu}}]+J[({\rho_J}+{p_J}){U^{'\mu}}{U^{'\nu}}+{p_J}{g^{\mu\nu}}],\eqno(20)
$$
where $\rho_C,p_C (\rho_C,p_C>0)$ and $\rho_J (\rho_J>0),p_J$ are
energy density and pressure respectively in real and imaginary
spacetime. Moreover $\rho_C,p_C $ are not variable with
$\rho_J,p_J$. We define
$$
s_{\mu\nu}{U^\mu}{U^\nu}=-1,~~~~~a_{\mu\nu}{U^{'\mu}}{U^{'\nu}}=-1,\eqno(21)
$$
$$
T_{\mu\nu}={g_{\mu\alpha}}{g_{\nu\beta}}{T^{\alpha\beta}},\eqno(22)
$$
From Eqs.(3),(20) and (22),we get
$$
T=(3{p_C}-{\rho_C})-{J^2}({\rho_J}+{p_J})+J(4p_J),\eqno(23)
$$
If we assume by separation of variables
$$
\mu(r,t)=\alpha(r,t)+J\beta(r,t)={a^2}(t)h(r)+J{a^2}(t)h(r),
$$
$$
\gamma(r,t)=\eta(r,t)+J\xi(r,t)={Y^2}(t){r^2}+J{Y^2}(t){r^2},\eqno(24)
$$
Substituting into Eqs.(14) and (19), we obtain a special solution
of equation $R_{01}=0$
$$
a(t)\approx Y(t),\eqno(25)
$$
Therefore, this gives rise to a metric of the form
$$
{dS^2}=-{dt^2}+{a^2}(t)[h(r){dr^2}+{r^2}({d\theta^2}+{\sin^2}\theta{d\phi^2})],\eqno(26)
$$
This is the cosmological solution in CSGT. For instance, complex
curvature $R_{00}$ and $R_{11}$ turn into
$$
R_{00}=3\frac{\ddot{a}}{a}
      =-4\pi G[(3{p_C}+{\rho_C})+{J^2}({p_J}-{\rho_J})]-J4\pi
      G[({p_C}-{\rho_C})+(2-{J^2}){\rho_J}+(4-{J^2}){p_J}],\eqno(27)
$$
$$
R_{11}=(a{\ddot{a}}h+2{\dot{a}^2}h)+\frac{h_r}{rh}+J(a{\ddot{a}}h+2{\dot{a}^2}h)
$$
$$
=4\pi G{a^2}h[({\rho_C}-{p_C})+{J^2}({\rho_J}-{p_J})]+J4\pi
G{a^2}h[({\rho_C}-{p_C})+{J^2}{\rho_J}+({J^2}-2){p_J}],\eqno(28)
$$
And by the conservation law of energy momentum
${T^{\mu\nu}}_{;~\nu}=0$, the following equation is obtained
$$
\dot{\rho_C}+\dot{p_C}+\frac{3\dot{a}}{a}(\rho_C+p_C)-\frac{\dot{p_C}}{1+J}+J[(\dot{\rho_J}+\dot{p_J})+\frac{3\dot{a}}{a}(\rho_J+p_J)]
-J\frac{\dot{p_J}}{1+J}=0,\eqno(29)
$$
where dot means derivative with respect to time.\\
Assuming $h(r)=1$, Eq.(26) is
$$
{dS^2}=-{dt^2}+{a^2}(t)[{dr^2}+{r^2}({d\theta^2}+{\sin^2}\theta{d\phi^2})],\eqno(30)
$$
The line element (30) is the spatially-flat FRW metric. And the
real and imaginary parts of complex curvature $R_{00}$ and
$R_{11}$ are respectively
$$ Q_{00}=\frac{3\ddot{a}}{a}=-4\pi
G[(3{p_C}+{\rho_C})+{J^2}({p_J}-{\rho_J})],\eqno(27a)
$$
$$
P_{00}=0=4\pi
G[({p_C}-{\rho_C})+(2-{J^2}){\rho_J}+(4-{J^2}){p_J}],\eqno(27b)
$$
and
$$
Q_{11}=2{\dot{a}^2}+a{\ddot{a}}=4\pi
G{a^2}[({\rho_C}-{p_C})+{J^2}({\rho_J}-{p_J})],\eqno(28a)
$$
$$
P_{11}=2{\dot{a}^2}+a{\ddot{a}}=4\pi
G{a^2}[({\rho_C}-{p_C})+{J^2}{\rho_J}+({J^2}-2){p_J}],\eqno(28b)
$$
By calculating, we obtain $J=\varepsilon$, i.e.the universe is the
hyperbolic complex symmetry space. For instance, Eqs.(27a) and
(28a) are
$$
\ddot{a}=-\frac{4\pi
G}{3}a[(3{p_C}+{\rho_C})+({p_\varepsilon}-{\rho_\varepsilon})],\eqno(31)
$$
$$
\frac{\dot{a}^2}{a^2}=\frac{8\pi
G}{3}[{\rho_C}+\frac{1}{2}({\rho_\varepsilon}-{p_\varepsilon})],\eqno(32)
$$
Eqs.(31) and (32) are modified Friedmann Equations, where
$\rho_\varepsilon$ and $p_\varepsilon$ are energy density and
pressure in hyperbolic imaginary space. Moreover, the relation of
$\rho_C,p_C$ and $\rho_\varepsilon,p_\varepsilon$ is
$$
{p_C}-{\rho_C}+{\rho_\varepsilon}+3{p_\varepsilon}=0,\eqno(33)
$$
\section{THE ACCELERATING EXPANSIBLE UNIVERSE}
In the section, we study the accelerating expansion of universe in
the hyperbolic complex symmetry gravitational theory (HCSGT). The
equation of state
$\omega_\varepsilon=\frac{p_\varepsilon}{\rho_\varepsilon}$.\\
\mbox{}\hspace{20pt}Eqs.(31) and (32) are rewritten as
$$
{H^2}=\frac{8\pi
G}{3}[{\rho_C}+\frac{1}{2}({\rho_\varepsilon}-{p_\varepsilon})],\eqno(34)
$$
$$
q
=-\frac{\ddot{a}a}{\dot{a}^2}=-\frac{\ddot{a}}{aH^2}=\frac{{\rho_C}+{3p_C}+({p_\varepsilon}-{\rho_\varepsilon})}{2{\rho_C}+({\rho_\varepsilon}-{p_\varepsilon})},\eqno(35)
$$
From Eq.(34), we get
$2{\rho_C}>{p_\varepsilon}-{\rho_\varepsilon}$ for $H^2>0$. Since
$q<0$, we obtain
$$
{\rho_C}+3{p_C}-{\rho_\varepsilon}+{p_\varepsilon}<0,\eqno(36)
$$
\mbox{}\hspace{20pt}Furthermore, we investigate the condition of
the accelerating expansion of universe if matter in the hyperbolic
imaginary space is taken as a new alternative of dark energy.\\
\mbox{}\hspace{20pt}If $p=0$, Eq.(33) is
$-\rho+3p_\varepsilon+\rho_\varepsilon=0$, i.e.
$\omega_\varepsilon>-\frac{1}{3}$. It satisfies the strong energy
condition (SEC) $\omega\geq -\frac{1}{3}$ [21]. But dark energy
must violate SEC in order to accelerate, i.e.
$\omega_\varepsilon<-\frac{1}{3}$. Substituting (33) into (36), we
obtain $\omega_\varepsilon>-\frac{1}{2}$.\\
\mbox{}\hspace{20pt}From the above analysis, we show if matter in
the hyperbolic imaginary space is taken as a new alternative of
dark energy in the HCSGT, the condition of the accelerating
expansion of universe is
$$
-\frac{1}{2}<\omega_\varepsilon<-\frac{1}{3},\eqno(37)
$$
The result is acceptable and consistent with ref [22].\\
\mbox{}\hspace{20pt}In the HCSGT, the conversion of energy
momentum (29) turn into
$$
\dot{\rho_C}+\frac{1}{2}\dot{p_C}+\frac{3\dot{a}}{a}({\rho_C}+{p_C})=0
$$
$$
\dot{\rho_J}+\frac{1}{2}\dot{p_J}+\frac{3\dot{a}}{a}({\rho_J}+{p_J})=0
$$
Substituting into Eq.(34), we get
$$
{H^2}=\frac{8\pi
G}{3}[{(1+z)^{3(1+\omega_C)}}{e^{-\frac{1}{2}\int\frac{d(\rho_C\omega_C)}{d\rho_C}}}+\frac{1}{2}{(1+z)^{3(1+\omega_\varepsilon)}}
(1-\omega_\varepsilon){e^{-\frac{1}{2}\int\frac{d(\rho_\varepsilon\omega_\varepsilon)}{d\rho_\varepsilon}}}],\eqno(38)
$$
\section{CONCLUSIONS}
In this paper we have proposed that CSGT may explain the
accelerating expansion of universe. We concretely take a real line
element and obtain the cosmological solution. Furthermore,
conditions of the accelerating expansion of universe are discussed
within HCSGT. Moreover, the equation of state of matter satisfies
$-\frac{1}{2}<\omega_\varepsilon<-\frac{1}{3}$
if the matter in the hyperbolic imaginary space is taken as a new alternative of dark energy.\\
\mbox{}\hspace{20pt}In above discussion, we have investigate the
conditions of the accelerating expansion of universe within CSGT.
But we don't deeply study the corresponding properties. These will
be explored in further work.
\section{ACKNOWLEDGEMENT}
We would like to thank Professor Ya-bo Wu and Doctor Li-xin Xu for
the helpful discussions. This work was supported by National
Science Foundation of China under Grant NO.10275008 and partly by
10475036.

\end{document}